\documentclass[
    ,final            
  ]
  {aipproc}

\layoutstyle{6x9}

\begin{document}

\title{EXACT STATIC AXIALLY SYMMETRIC THIN ANNULAR DUST DISKS}


\author{Guillermo A. Gonz\'{a}lez}{
  address={Escuela de F\'isica -- Universidad Industrial de Santander \\
A. A. 678, Bucaramanga, Colombia},
  altaddress={Departamento de F\'isica Te\'orica -- Universidad del Pa\'is Vasco \\
48080, Bilbao, Espa\~na} 
}

\author{Antonio C. Guti\'errez-Pi\~{n}eres}{
  address={Escuela de F\'isica -- Universidad Industrial de Santander \\
A. A. 678, Bucaramanga, Colombia}
}

\author{Viviana M. Vi\~{n}a-Cervantes}{
  address={Escuela de F\'isica -- Universidad Industrial de Santander \\
A. A. 678, Bucaramanga, Colombia}
}

\begin{abstract}
A new family of exact solutions of the Einstein field equations for static and
axially simmetric spacetimes is presented. All the metric functions of the
solutions are explicitly computed and the obtained expressions are simply
written in terms of oblate spheroidal coordinates. The solutions describe an
infinite family of static axially symmetric thin annular dust disks. The disk
are of infinite extension but with an inner annular edge. The energy densities
of the disks are everywhere positive functions of the radial coordinate, equals
to zero at the inner edge of the disk, having a maximun at a finite value of the
radius and then vanishing at infinity. The disks have finite mass and their
energy-momentum tensor agrees with all the energy conditions.
\end{abstract}

\maketitle


The Weyl metric for a static axially symmetric spacetime is \cite{KHSM}
\begin{equation}
ds^2 = - e^{2\Phi}dt^2 + e^{-2\Phi}[r^2d\varphi^2 + e^{2\Lambda}(dr^2 + dz^2)],
\label{eq:met}
\end{equation}
with $\Phi$ and $\Lambda$ only depending on $r$ and $z$. The Einstein vacuum
equations are
\begin{eqnarray}
&&\Phi_{,rr} + \frac{1}{r} \Phi_{,r} + \Phi_{,zz} \ = \ 0 \ , \label{eq:weyl1} 
\\
&&\Lambda_{,r} \ = \ r ( \Phi_{,r}^2 - \Phi_{,z}^2 ) \ , \label{eq:weyl2}   \\
&&\Lambda_{,z} \ = \ 2 r \Phi_{,z} \Phi_{,z} \ . \label{eq:weyl3}
\end{eqnarray}\label{eq:weyeq}
We now impose the conditions
\begin{eqnarray}
\Phi(r,z) &=& \Phi(r,-z),\label{eq:con1}
\end{eqnarray}
and
\begin{eqnarray}
\Phi_{,z}(r,0^{+}) &=& \left\{ \begin{array}{ll}
0 &; \;\; 0 \leq r \leq a, \\
f(r) & ; \;\;r \geq a. \\
\end{array}\label{eq:con2}\right. 
\end{eqnarray}
In order to solve the Einstein equations, we introduce the oblate spheroidal
coordinates by means of
\begin{equation}
r^2 = a^2 (x^2 + 1)(1 - y^2), \qquad z = a x y,
\end{equation}
with $- \infty < x < \infty$, $0 \leq y \leq 1$. The disk is obtained by taking
$y = 0$ and so is located at $z = 0$, $r \geq a$. On crossing the disk, $x$
changes sign but does not change in absolute value, so that an even function of
$x$ is a continuous function everywhere but has a discontinuous $x$ derivative
at the disk.

A conveniente solution, satisfying (\ref{eq:con1}) and (\ref{eq:con2}) and
regular for $y \neq 1$, is given by \cite{GG1}
\begin{eqnarray}
\Phi_{_{0}}(x,y)=
\frac{\alpha}{2}\ln\left[\frac{1 + y}{1 -y}\right]\label{eq:phidef},
\end{eqnarray}
\begin{eqnarray}
\Lambda_{_{0}}(x,y) = 
\frac{\alpha^2}{2}\ln\left[\frac{1 - y^2}{x^2 + y^2}\right],
\end{eqnarray}
where  $\alpha$ is an arbitrary constant and $a$ the inner radius of the disk.
So, by using the distributional approach \cite{PH, LICH, TAUB}, the Surface
Energy-Momentum Tensor of the disk can be written as
\begin{equation}
S_{ab} \ = \ \epsilon V_a V_b , \label{eq:emt0}
\end{equation}
where $V^a = e^{- \Phi} \delta_{_0}^a$. The surface energy density is given by
\begin{equation}
\epsilon = \frac{4 \alpha}{a} x^{\alpha^2 - 1}. 
\end{equation}
Note that, if $\alpha > 0$ then $\epsilon > 0$, i.e. the disk satisfy all the
energy conditions. Nevertheless, for $\alpha^2 > 1$ we have that $\epsilon$
increases to infinite when $r \rightarrow \infty$, whereas for $\alpha^2 <1$
$\epsilon$ increases to infinite at the edge of the disk, $r=a$.

From (\ref{eq:phidef}) we can generate an infinite family of solutions by means
of the procedure  \cite{GG1}
\begin{eqnarray}
\Phi_{n+1}(x,y;a) =  \frac{\partial \Phi_{_{n}}(x,y;a)}{\partial
a}\label{eq:algorithm}
\end{eqnarray}
where $n \geq 0$. The first member of the new family of solutions is
\begin{eqnarray}
\Phi_{1}(x,y;a) &=& \frac{\alpha y}{a(x^2 + y^2)}, \label{eq:phi1edge} \\
\Lambda_{1}(x,y;a) &=& \frac{\alpha^2(y^2 - 1) A_1(x,y)}{4a^2(x^2 + y^2)^4},
\end{eqnarray}
with
$$
A_1(x,y) = 9x^4y^2 - x^4 +2x^2y^4 + 6x^2y^2 + y^6 - y^4,
$$
and the second solution of the family is 
\begin{eqnarray}
\Phi_{2}(x,y;a) & =&\frac{\alpha y }{a^2(x^2 + y^2)^3}\{x^4 - 3x^2y^2 + 3x^2 -
y^2\},\\
\Lambda_{2}(x,y;a) &=& \frac{\alpha^2(y^2 - 1) A_2(x,y)}{8a^4(x^2 + y^2)^8},
\end{eqnarray}
with
\begin{eqnarray*}
A_2(x,y) &=& 2x^{12}(9y^2 -1 ) - 4x^{10}(51y^4 -41y^2 + 2) \\
&&+ x^8(735y^6 -1241y^4 + 419y^2 -9) \\
&&- x^6y^2(132y^6 -1644y^4+ 1604y^2 -252) \\
&&+ x^4y^4(84y^6 -384y^4 + 1266y^2 -630) \\
&&+ 4x^2y^6(6y^6 +6y^4 -39y^2 + 63) + 3y^8(y^6 + y^4 + y^2 -3).
\end{eqnarray*}
For $n \geq 3$ the solutions are more involved, but can also easily obtained.

The surface energy densities of these two thin disk models are
\begin{eqnarray}
\epsilon_1 &=& \frac{4\alpha}{a^2x^3} e^{-(\alpha^2/4a^2x^4)}, \\
\epsilon_2 &=& \frac{4\alpha(x^2 + 3)}{a^3x^5} e^{-\alpha^2(2x^4 +8x^2
+9)/(8a^4x^8)}.
\end{eqnarray}
for $x \geq 0$, and will be allways positive if we take $\alpha > 0$. We then
have dust disks in agreement with all the energy conditions. The total mass of the disks can be easily computed and, for the first
solution, we obtain
\begin{equation}
M_1 = 2 \pi \ \Gamma (1/4) \sqrt{2 a \alpha} \ ,
\end{equation}
with similar expressions for all the members of the family. So, these solutions
describe an infinite family of thin dust disks with a central inner edge, whose
energy densities are everywhere positive and well behaved, in such a way that
their energy-momentum tensor are in fully agreement with all the energy
conditions. Moreover, although the disks are of infinite extension, all of them
have finite mass.

\begin{figure}
  \includegraphics[height=.3\textheight]{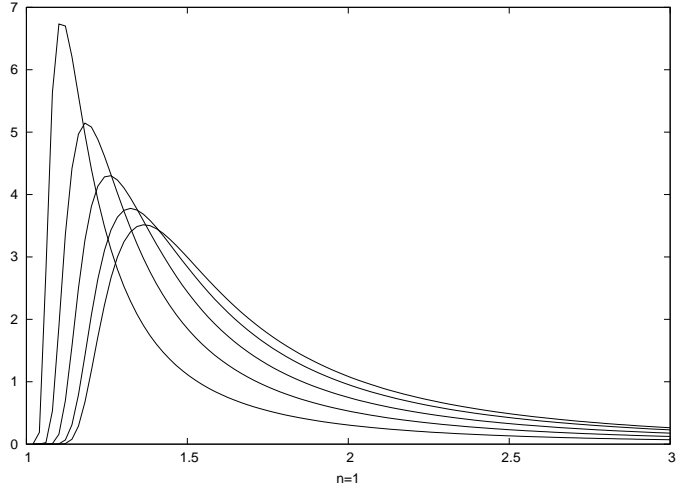} 
\caption{${\tilde \epsilon} = a \epsilon$ as a function of ${\tilde r} = r/a$ 
for the disk with  $n= 1$ and $\alpha$ = 0.4, 0.7, 1, 1.3 and 1.5.}
\end{figure}
\begin{figure}
  \includegraphics[height=.3\textheight]{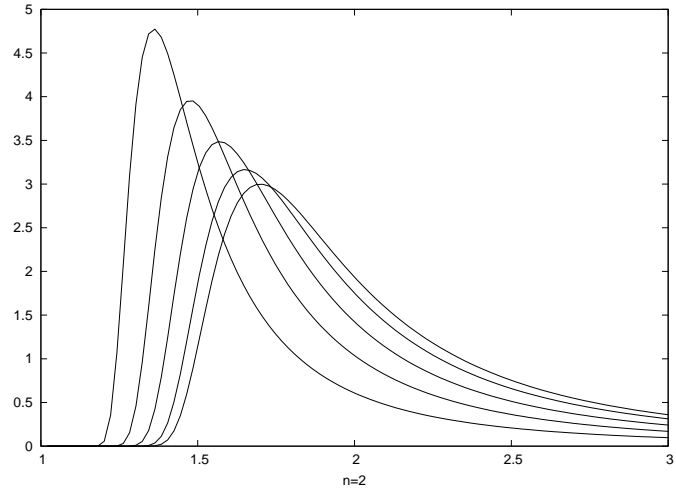}
\caption{${\tilde \epsilon} = a \epsilon$ as a function of ${\tilde r} = r/a$ 
for the disk with $n= 2$ and $\alpha$ = 0.4, 0.7, 1, 1.3 and 1.5.}
\end{figure}

Now, as all the metric functions of the solutions are explicitly computed, these
are the first fully integrated explicit exact solutions for such kind of thin
disk sources. Furthermore, their relative simplicity when expressed in terms of
oblate spheroidal coordinates, makes it very easy to study different dynamical
aspects, like the motion of particles inside and outside the disks and the
stability of the orbits. Now, besides their importance as a new family of exact
and explicit solutions of the Einstein vacuum equations, the main importance of
this family of solutions is that they can be easily superposed with the
Schwarzschild solution in order to describe binary systems composed by a thin
disk surrounding a central black hole \cite{GG2}.


\begin{theacknowledgments}
A. C.~Guti\'errez-Pi\~{n}eres wants to thank the financial support from
COLCIENCIAS, Colombia.
\end{theacknowledgments}



\bibliographystyle{aipproc}   


\begin{thebibliography}{9999}

\bibitem{KHSM} H. Stephani, D. Kramer, M. MacCallum, C. Hoenselaers and E.
Herlt, {\it Exact Solutions of Einstein's Field Equations}. (Cambridge
University Press, 2003).

\bibitem{GG1} G. A. Gonz\'alez, A. C. Guti\'errez-Pi\~neres and V. M.
Vi\~na-Cervantes, {\it Relativistic static thin dust disks with an inner edge: 
An infinite family of new exact solutions}, arXiv: 0811.3869 (2008).

\bibitem{PH} A. Papapetrou and A. Hamouni, Ann. Inst. Henri Poincar\'e {\bf 9},
179 (1968).

\bibitem{LICH} A. Lichnerowicz, C.R. Acad. Sci. {\bf 273}, 528 (1971).

\bibitem{TAUB} A. H. Taub, J. Math. Phys. {\bf 21}, 1423 (1980).

\bibitem{GG2} G. A. Gonz\'alez and A. C. Guti\'errez-Pi\~neres, {\it
Counterrotating Dust Disk Around a Schwarzschild Black Hole: New Fully
Integrated Explicit Exact Solution}, arXiv: 0811.3002v1 (2008).


\end{thebibliography}

\end{document}